\documentclass[aps,prb,twocolumn,showpacs,preprintnumbers, superscriptaddress]{revtex4}

\usepackage{graphicx}

\begin{document}


\title{ Stretched  exponential  behavior in  remanent lattice striction

of a (La,Pr)$_{1.2}$Sr$_{1.8}$Mn$_{2}$O$_{7}$ bilayer manganite single crystal}

\author{M.Matsukawa} 
\email{matsukawa@iwate-u.ac.jp }
\author{M.Chiba}
\affiliation{Department of Materials Science and Technology, Iwate University , Morioka 020-8551 , Japan }
\author{R.Suryanarayanan}
\author{M.Apostu}
\author{A.Revcolevschi}
\affiliation{Laboratoire de Physico-Chimie de L'Etat Solide,CNRS,UMR8648
 Universite Paris-Sud, 91405 Orsay,France}
\author{}
\affiliation{}
\author{S.Nimori}
\affiliation{National Institute for Materials Science, Tsukuba 305-0047 ,Japan}
\author{ N. Kobayashi }
\affiliation{ Institute for Materials Research, Tohoku University, Sendai  980-8577, Japan }

\date{\today}

\begin{abstract}
We have investigated the time dependence of remanent magnetostriction in a (La,Pr)$_{1.2}$Sr$_{1.8}$Mn$_{2}$O$_{7}$ single crystal, 
in order to examine the slow dynamics of  lattice distortion in bilayered manganites.
A competition between double exchange and Jahn-Teller type orbital-lattice interactions results in the observed lattice profile following  a  stretched  exponential  function. 
This finding  suggests that spatial growth of  the local lattice distortions coupled with e$_{g}$-electron orbital strongly correlates with the appearance of the field-induced CMR effect.  

\end{abstract}

\pacs{75.47.Lx,75.50.Lk}

\maketitle

The discovery of colossal magnetoresistance (CMR)  effect in doped manganites with perovskite 
structure has stimulated considerable interest for the understanding of their physical properties \cite{TO00}. 
Though the insulator to metal (IM) transition and its associated CMR are well explained  
on the basis of  the double exchange (DE) model,  it is pointed out that the dynamic John-Teller 
(JT) effect due to the strong electron-phonon interaction, plays a significant role in 
the appearance of CMR as well as the DE interaction \cite{ZE51,MI95}.  Furthermore, Dagotto 
et al propose a phase separation model where  the ferromagnetic (FM) 
metallic and antiferromagnetic (AFM) insulating clusters coexist and their model strongly
 supports recent experimental studies on the physics of manganites  \cite{DA01}. 
 
 The bilayer manganite La$_{1.2}$Sr$_{1.8}$Mn$_{2}$O$_{7}$ exhibits a paramagnetic insulator (PMI) to 
 ferromagnetic metal (FMM) transition around  $T_{c}=\sim$120K  and  its  associated CMR effect \cite{MO96}. 
In comparison with cubic manganites, the MR effect of the compound under consideration, due to its layered structure,
is enhanced by  two orders of magnitude, at 8T, around $T_{c}$. 
It is well known that Pr-substitution on the La-site leading to (La$_{1-z}$,Pr$_{z}$)$_{1.2}$Sr$_{1.8}$Mn$_{2}$O$_{7}$ 
causes an elongation of the $c$ axis length in contrast with  a shrinkage of the $a$($b$) axis, resulting 
in a change of the e$_{g}$-electron occupation from the d$_{x^2-y^2}$ to d$_{3z^2-r^2}$ orbital \cite{MO97,OG00}. 
 For $z$=0.6, a strong suppression of the e$_{g}$-electron bandwidth 
leads to  a PM insulating ground state \cite{AP01}. However,  the FM metastable 
phase with a metallic conduction is easily obtained by the application of a magnetic field.   
 The field induced first-order IM transition is further
accompanied by a memory effect of magnetoresistance, magnetization, magnetostriction and 
magnetothermal conductivity \cite{AP01,GO01,MA03}.  This memory effect may be explained  
by a schematic picture of the free energy exhibiting two local minima corresponding to the FMM and PMI 
states, as shown in the inset of Fig.1(a). In the absence of magnetic field, the system stays in the stable PI state but, upon field application, the FMM state becomes metastable against the PMI state.  After removing the field, the system should go back to a pure PMI ground state through the various kinds of mixed states after a very long time. 
Recently,  Gordon et al \cite{GO01}  have reported that  the  time  dependence  of  remanent  magnetization of (La,Pr)$_{1.2}$Sr$_{1.8}$Mn$_{2}$O$_{7}$ follows  a stretched exponential function, which is closely related to a magnetic frustration between FM double exchange and AFM super exchange interactions. 
In several reports concerning cubic manganites,  the magnetization and resistivity relaxation effects have been discussed
 within the framework of the phase segregation between FM metal and AFM charge ordered insulating states 
\cite{AN99,SM99,UE00,LO01}.  In addition, there are a few reports on magnetic and transport properties 
in doped bilayer manganites which extend our understanding of a spin-glass like phase related to  the phase 
separation model \cite{DH01,GO01}. However, to our knowledge, slow dynamics of a lattice relaxation 
in doped manganites has not been reported so far. 
       
In this paper, we report  the  time dependence of remanent magnetostriction in the  
Pr-substituted bilayer manganites. A frustration of lattice deformation between double exchange and Jahn-Teller type orbital-lattice interactions plays a crucial role in the observed slow relaxation going from the FM metal to PM insulator.


Single crystals of (La$_{0.4}$,Pr$_{0.6}$)$_{1.2}$Sr$_{1.8}$Mn$_{2}$O$_{7}$ were grown by the floating zone method
 using  a mirror furnace. 
The calculated lattice parameters were reported earlier in \cite{AP01}.  The dimensions of  the $z$=0.6 sample are 
3.4$\times$3 mm$^2$ 
in the $ab$-plane and 1mm along the $c$-axis.  Magnetostriction, both in the $ab$-plane and along the $c$-axis ,
 was measured by means of a conventional strain gauge method at the Tsukuba Magnet Laboratory, 
the National Institute for Materials Science and at the High Field Laboratory for Superconducting Materials, Institute
 for Materials Research, Tohoku University. The magnetization measurements were made  by using a superconducting quantum
 interference device  magnetometer at Iwate University.  A magnetic phase diagram in the ($H,T$) plane was established 
from the magnetic measurements on the $z$=0.6 crystal as shown in the inset of Fig1.(b) \cite{AP01,APDC}.  
The  phase  diagram  is  separated  into  three  regions labeled as  the PMI, FMM and mixed phases 
(the hatched area in the ($H,T$) plane). In the mixed phase, the PMI and FMM clusters coexist, which is characterized 
by  hysteresis in the magnetization curves. Thus, we tried to look for a  relaxation effect for  the $z$=0.6 crystal  
at  temperatures selected between 20 and 30K, covering a  large to small hysterises region. 
After cooling the sample  down  to  the respective temperatures in a zero field, the field was applied along the c-axis 
up to 5T at a sweep rate of  0.2 T/min and was then decreased down to zero at the same rate. A dotted horizontal line 
in the phase diagram
 represents such an experimental procedure crossing the phase boundary separating the hysteresis and FM areas. 
Finally, magnetostriction data were recorded as a function of time just after the field was switched off.    
\begin{figure}[ht]
\includegraphics[width=8cm]{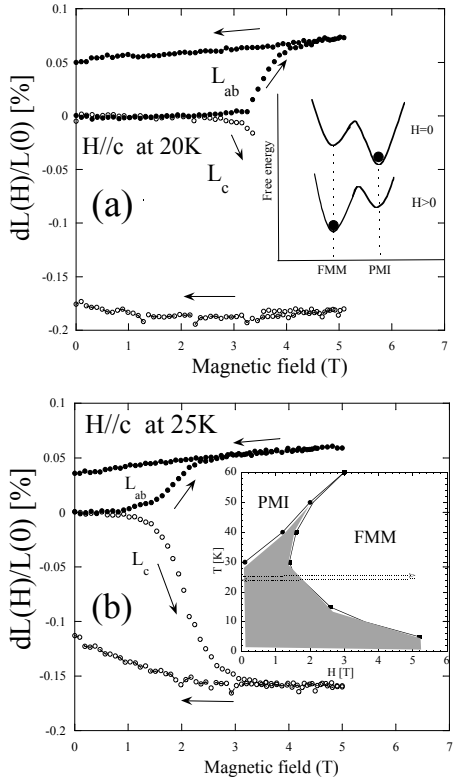}%
\caption{Magnetostriction both in the $ab$-plane and along  the $c$-axis, 
for  single crystalline 
(La$_{0.4}$,Pr$_{0.6}$)$_{1.2}$Sr$_{1.8}$Mn$_{2}$O$_{7}$. A schematic picture of the free energy 
with two local minima corresponding to the FMM and PMI 
states is shown in the inset of Fig.1(a). The inset of Fig.1(b) represents  a magnetic phase diagram
in the ($H,T$) plane established from the magnetic measurements carried out on the $z$=0.6 crystal.}
\end{figure}%

\begin{figure}[ht]
\includegraphics[width=8cm]{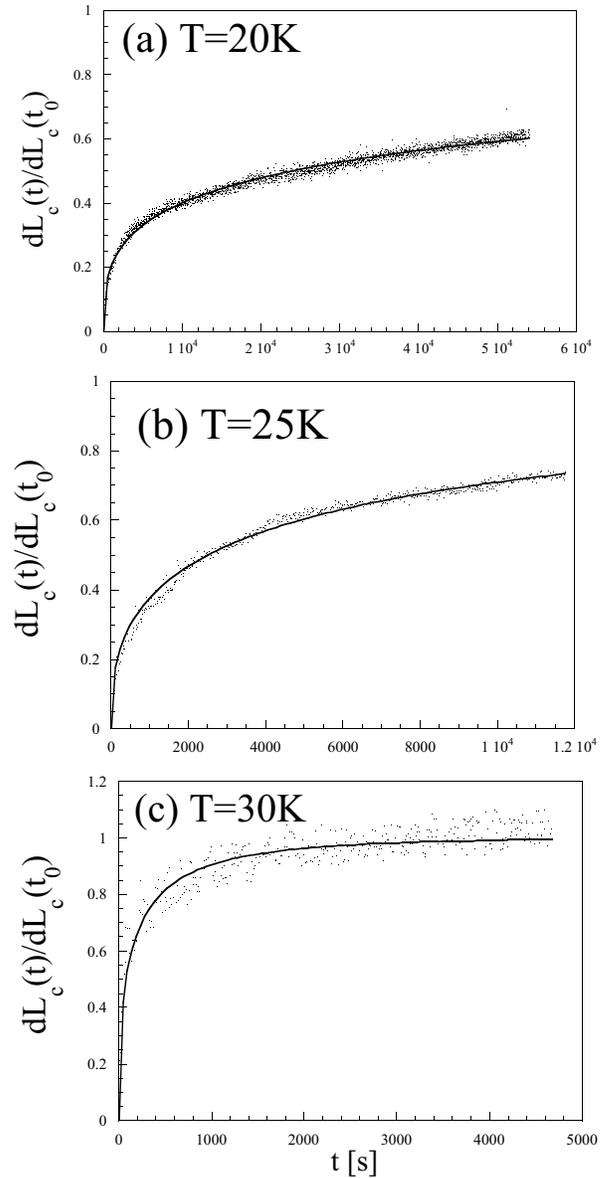}%
\caption{The remanent $c$-axis normalized magnetostriction  of single crystalline (La$_{0.4}$,Pr$_{0.6}$)$_{1.2}$Sr$_{1.8}$Mn$_{2}$O$_{7}$ 
as a function of time, just after the field was switched off. The solid lines 
correspond to  a fit by  a stretched exponential function , $dL_{c}(t)/dL_{c}(t_{0})$=1-exp$[-(t/\tau)^\beta]$, where  $\tau$ and $\beta$ represent the  characteristic relaxation  time  and exponent, respectively. (a)$\tau$=$6.9\times10^4$ s and $\beta$=0.35 at 20K,(b)$\tau$=$6.0\times10^3$ s and $\beta$=0.42 at 25K and (c)$\tau$=$1.7\times10^2$ s and $\beta$=0.48 at 30K. } 
\end{figure}%

Magnetostriction data both in the ab-plane and along  the c-axis , 
for the (La$_{0.4}$,Pr$_{0.6}$)$_{1.2}$Sr$_{1.8}$Mn$_{2}$O$_{7}$ single crystal are shown in Fig.1 as a function of magnetic field applied parallel to the $c$-axis at selected temperatures.  At 20K, 
 the value of  $dL_{c}(H)/L_{c}(0)$ exhibits an abrupt decrease  around  3T , while the value of $dL_{ab}(H) /L_{ab}(0)$ rises
at the same field value. Here, the value of $dL_{i}(H)$ is defined as $L_{i}(H)-L_{i}(0)$. 
In other words,  the $c$-axis length suddenly shrinks around 3T in contrast with the expansion of the $a$($b$) 
axis accompanied by the field-induced IM transition.  Combining $dL_{ab}(H) /L_{ab}(0)$ and $dL_{c}(H)/L_{c}(0)$,
 a volume shrinkage associated with the IM transition is deduced, indicating a drastic suppression of the local lattice 
distortion associated with the transition from the localized 
to itinerant  state \cite{IB95}. 
The resultant volume striction remains invariant down to zero field  although  its magnitude depends on 
the respective temperatures.   
In the insulating state of hole doped bilayer manganites,  there exists a lattice disorder
 between the  Mn$^{3+}$O$_{6}$ and Mn$^{4+}$O$_{6}$  
octahedral, with and without the JT local lattice distortion, respectively\cite{ME99}. 
 As we pass $T_{c}$ from the insulating state, 
the Mn-O bond length disorder  is increasingly  suppressed and, in the metallic state, it becomes less inhomogeneous than 
that found in the insulating state. Giant magnetostriction of  the $z$=0.6 crystal is understood on the basis of a strong  suppression of the local lattice disorder due to the field-induced FM phase, where the double exchange interaction is favorable against the orbital-lattice interaction.
 Moreover, the polarized  neutron diffraction study on the z=0.6 crystal 
suggests that the field-induced metallic state, where the field is parallel to the $c$-axis, is characterized by a high population of 
the d$_{3z^2-r^2}$ orbital of Mn$^{3+}$ \cite{WA03}. This finding is consistent with the $c$-axis distortion based on the local JT effect in the insulating state, preferring occupation of the d$_{3z^2- r^2}$ orbital through the orbital-lattice coupling.        

Next, let us show in Fig.2 the remanent $c$-axis magnetostriction data of the (La$_{0.4}$,Pr$_{0.6}$)$_{1.2}$Sr$_{1.8}$Mn$_{2}$O$_{7}$ 
single crystal 
as a function of time, just after the field was switched off . Here, the value of  $dL_{c}(t)/dL_{c}(t_{0})$ denotes 
the normalized 
$dL_{c}(t)=[L_{c}(t)-L_{c}(0)]$  by $dL_{c}(t_{0})=[L_{c}(t_{0}\ll0)-L_{c}(0)]$  where $L_{c}(t_{0}\ll0)$ 
and $L_{c}(t=0)$correspond to the virgin and initial values, 
before the application of field and just after the removal of field, respectively.  If  the system fully relaxes to the PM virgin state ,
 $dL_{c}(t)$ should approach an equilibrium value of $dL_{c}(t_{0})$. It should be noted that, at higher temperatures, 
above 30K, 
 the remanent striction instantaneously decays just after removing field,  while at low temperatures (5K),
 the value of $dL_{c}(t)$  
remains constant for 1 day.  At 30K, the lattice striction rapidly rises after removal of the field,
then it  relaxes 
by a half of the total value up to the relaxation time (=$1.7\times10^2$ s) and 
finally restores the ground-state value. 
Furthermore, at lower temperatures, it takes a much longer time for $dL_{c}$  to decay to the ground-state value, 
as shown in Fig.2. 
 The temporal profile of the remanent magnetostriction follows a stretched exponential function 
$dL_{c}(t)/dL_{c}(t_{0})$=1-exp$[-(t/\tau)^\beta]$, where  $\tau$ and $\beta$ represent the  characteristic 
 relaxation  time  and exponent, respectively. 
The fitted parameters, $\tau$ and $\beta$, are plotted as a function of  temperature in Fig.3(a). 
Here, the temperature dependence of $\tau$  is well described by a thermally activated function
$\tau=A$exp($\Delta /kT)$
where $A=2.0\times10^{-2}$ and $\Delta =30$meV. Moreover, the value of the exponent is roughly fitted 
by  a linear function of the form  $\beta =3\times10^{-2}+T/66.7$. Such a deviation from the simple exponential decay needs a further treatment 
on the basis of a multiple relaxation time approximation to describe  the observed  relaxation effect \cite{PA84}.  
The transition probability ($\tau^{-1}$) from the FMM metastable to the PMI stable states is proportional to 
exp(-$\delta U/kT$), where $\delta U$ represents potential barriers between the metastable state and the local maximum. 
A  wide  distribution  of  the potential  barriers is related to a stretched exponential decay, leading to our understanding of the phase separation between FM and PM clusters.  
\begin{figure}[ht]
\includegraphics[width=8cm]{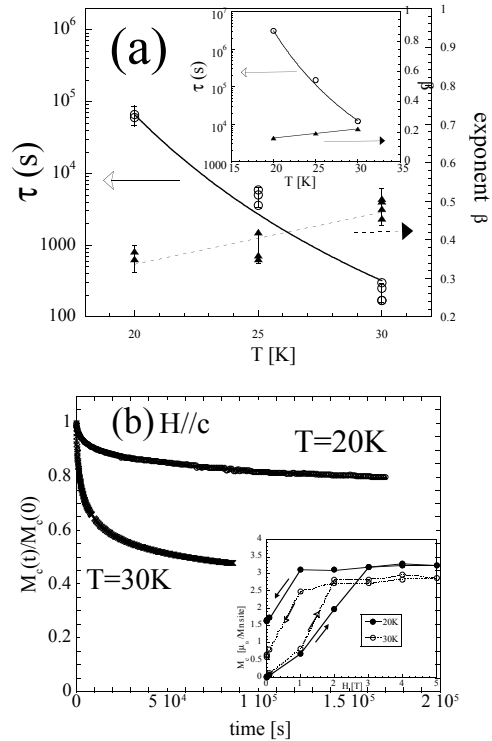}
\caption{(a) remanent magnetostriction.  Characteristic relaxation  time  and exponent, $\tau$ and $\beta$, respectively, plotted as a function of temperature. The former is well fitted by a thermally activated function, while the latter follows a $T$-linear function. For comparison, in the inset of Fig.3(a), the fitted parameters to remanent magnetization data are also shown. 
(b) $c$-axis  normalized  magnetization of  the same crystal as a function of time after removing 
the field. $\tau$=$3.1\times10^6$ s and $\beta$=0.16 at 20K, 
$\tau$=$1.2\times10^4$ s and $\beta$=0.22 at 30K. In the inset of Fig.3(b), the  data collected for $M_{c}(H)$ before measuring the remanent magnetization, are also shown. }
\end{figure}%

For comparison, the $c$-axis magnetization of the same crystal is presented in Fig.3(b) as a function of time after removing the field.  In the inset of Fig.3(b), are also shown the data collected for $M_{c}(H)$ before measuring the remanent magnetization. 
In  the inset of Fig.3(a),  the parameters, $\tau$ and $\beta$, obtained by fitting the magnetization data, are plotted as a function of temperature\cite{COM1}.  In a similar way, the activation energy of the relaxation time is estimated to be 33 meV and the exponent is fitted to the $T$-linear function$\beta =4\times10^{-2}+T/167$ . We notice that the characteristic $\tau$ of $L_{c}$ is about by two orders of magnitude 
smaller than that of $M_{c}$ on the same crystal, accompanied by the higher exponent .   

Here, we comment some differences in the relaxation processes between the magnetostriction and magnetization although both decay curves follow a similar stretched exponential function.  
First, in the view point of a frustration among the competing interactions, both the situations are different. The remanent magnetostriction slowly relaxes because of  the competition between the double exchange and JT type lattice-orbital interactions.
The former interaction causes a suppression in the local lattice distortion along the $c$-axis through the itinerant state, but the latter favors the lattice deformation through the local John-Teller effect. On the other hand,
in a previous study carried out on the $z$=0.6 crystal,  the temperature variation of the magnetization   exhibited  a  steep decrease in the ZFC scan below  $T_{g}$=28K, which is probably associated with a spin-glass state \cite{AP01}. 
It is thus expected from the spin-glass like behavior of $ M(T)$  that the decay process of remanent magnetization is closely related to a magnetic frustration between FM double exchange and AFM super exchange interactions. 
Furthermore, we point out the discrepancy in the fitted relaxation time between $L(t)$ and $M(t)$ curves as shown in Fig.3.  A gradual drop in $M(H)$ curve at 20K with decreasing field is quite contrast with field-independent  behaviors of $L(H)$ and $\rho(H)$ at low temperatures during the field sweep to zero \cite{GO01}.  The similarity in remanent values between $L(H)$ and $\rho(H)$ data strongly suggests that the system still remains a metallic state with a volume shrinkage just after removing the field. As one of reasons for the difference between $M(H)$ and $L(H)$ (or $\rho(H)$), we think that  the nucleation of magnetic domains in the metallic region probably gives a decrease in $M(H)$ at low fields. 

Recently, Y.Tokunaga et al. have revealed the existence of magnetic domain structures in (La$_{0.4}$,Pr$_{0.6}$)$_{1.2}$Sr$_{1.8}$Mn$_{2}$O$_{7}$ 
single crystal using a magneto-optical imaging technique \cite{TO04}.
It is true that the formation of domain walls never produce a much longer time of magnetic relaxation than the lattice relaxation. However,  some of domain walls and/or each spins are pinned at lattice defect sites, giving a long lifetime of magnetic domains. Another scenario  for the disagreement is concerning the existence of  magnetic polarons in the insulating state, causing a magnetic contribution. In particular,  in the hole doped bilayer manganites, Zhou et al., proposed that  Zener polarons occupying two Mn sites forms a ferromagnetic cluster in the paramagnetic regime   above $T_{c}$, to explain the pressure effect on thermoelectric power \cite{ZH98}.       
%
%
In addition, it should be noted that  the  Debye-Waller (DW) factor  
along the Mn-apical oxygen bond in the insulating state is greater \cite{ME99} than  the DW in the metallic one, at most by a factor of 2. In other words,  the difference in the local lattice inhomogeneity between the metallic and insulating states is not as large as that in $M$ between the ferromagnetic and paramagnetic states.
Finally, we make some comments on the relaxation phenomena of remanent resistivity in manganites. The resistive relaxation studies support that the insulator-metal transition and its associated CMR effect are well described  on the view point of both the  phase separation and  its related percolation behavior of metallic phase \cite{AN99,CH03}. The spatial growth of  the local lattice distortions coupled with e$_{g}$-electron orbital strongly correlates with the appearance of the field-induced CMR effect through the spatial evolution of  percolating paths. 

In summary, we have reported  the  time dependence of remanent magnetostriction in the  Pr-substituted bilayer manganites. A frustration between double exchange and Jahn-Teller type orbital-lattice interactions plays a crucial role on the observed slow  relaxation of lattice  going from the field-induced FMM to PMI states. 

This work was partially supported by a Grant-in-Aid for Scientific Research from the Ministry of Education, Science and Culture, Japan.



%


\end{document}